\documentclass[reprint,pre]{revtex4-1}

\usepackage{amsmath} 
\usepackage{amsthm} 
\usepackage{amssymb}	
\usepackage[pdftex]{graphicx}
\usepackage{epstopdf}
\usepackage{enumerate}
\usepackage{wrapfig}
\usepackage[dvips,letterpaper,margin=1in,bottom=1in]{geometry}
\usepackage{color}
\usepackage{tensind}
\tensordelimiter{`}
\whenindex{a}{\alpha}{} \whenindex{b}{\beta}{} \whenindex{g}{\gamma}{} \whenindex{d}{\delta}{} \whenindex{e}{\varepsilon}{} \whenindex{m}{\mu}{} \whenindex{n}{\nu}{} \whenindex{l}{\lambda}{} \whenindex{k}{\kappa}{} \whenindex{r}{\rho}{} \whenindex{s}{\sigma}{} \whenindex{t}{\tau}{}
\usepackage{todonotes}
\usepackage{subcaption}
\usepackage{setspace}
\usepackage{comment}
\usepackage{cancel}
\usepackage[]{units}
\usepackage{floatrow}

\usepackage{subcaption}
\usepackage{ragged2e}
\DeclareCaptionJustification{justified}{\justifying}
\usepackage{hyperref}
\usepackage[all]{hypcap}
\makeatletter 

\renewcommand{\v}[1]{\ensuremath{\mbox{\boldmath$ #1 $}}} 
 
\newcommand{\pd}[2]{\frac{\partial #1}{\partial #2}} 
 
\let\baraccent=\= 
\renewcommand{\=}[1]{\stackrel{#1}{=}} 
\renewcommand{\(}{\left (}
\renewcommand{\)}{\right  )}
\renewcommand{\[}{\left [}
\renewcommand{\]}{\right ]}
\newcommand{\<}{\left <}
\renewcommand{\>}{\right >}

\theoremstyle{definition}

\theoremstyle{remark}

\newtheorem{?}{\textbf{Question}}

\newcommand{\e}[1]{\mbox{e}^{#1}} 
\renewcommand{\exp}[1]{\mbox{exp}\[#1\]} 

\newcommand{\bigO}[1]{O\(#1\)}  

\begin{document}
\title{Beyond RG: from parameter flow to metric flow}
\begin{abstract}
Complex systems with many degrees of freedom are typically intractable, but some of their behaviors may admit simpler effective descriptions. The question of when such effective descriptions are possible remains open. The paradigmatic approach where such ``emergent simplicity'' can be understood in detail is the renormalization group (RG). Here, we show that for general systems, without the self-similarity symmetry required by the RG construction, the RG flow of model parameters is replaced by a more general flow of the Fisher Information Metric on the model manifold. We demonstrate that the systems traditionally studied with RG comprise special cases where this metric flow can be induced by a parameter flow, keeping the global geometry of the model-manifold fixed. In general, however, the geometry may deform, and metric flow cannot be reduced to a parameter flow -- though this could be achieved at the cost of augmenting the manifold by one new parameter, as we discuss. We hope that our framework can clarify how ideas from RG may apply in a broader class of complex systems.
\end{abstract}

\author{Charlotte Strandkvist}
\thanks{These authors contributed equally to the work}
\affiliation{Department of Systems Biology, Harvard University, 200 Longwood Avenue, Boston, MA 02115, USA}
\author{Pavel Chvykov$^*$}
\affiliation{Physics of Living Systems, Massachusetts Institute of Technology, Cambridge, MA 02139, USA}
\author{Mikhail Tikhonov}
\affiliation{Department of Physics; Center for Science and Engineering of Living Systems, Washington University in St.~Louis, St.~Louis, MO 63130, USA.}

\maketitle

The macroscopic behavior of many physical systems can be described by effective theories that are largely independent of the underlying microscopic degrees of freedom. For example, continuum mechanics makes no reference to the atomic basis of matter and statistical physics was conceived without knowledge of elementary particle physics. As noted by Goldenfeld and Kadanoff, this hierarchical character is the reason we ``don't model bulldozers with quarks''~\cite{goldenfeld1999}. 

The canonical construction for coarse-graining in physics is the renormalization group (RG). This collection of ideas describes how to systematically eliminate degrees of freedom, formalizing the construction of physical theories at different scales. The renormalization group establishes a hierarchy of length/energy scales and describes how theories transform as the scale of observation grows. Upon coarse-graining, only a few parameter combinations end up being relevant and these determine the effective theory. The renormalization group is one of the conceptually most profound tools in theoretical physics; being ``a theory about theories" it has enabled physicists to formally study how effective theories emerge~\cite{goldenfeld2018lectures}, and linked the notion of an effective theory to the notion of a hierarchy of scales. 

However, the RG construction relies on the presence of special symmetries (i.e., self-similarity across scales) that in many systems, for example in biology, have no sensible counterpart, thus precluding model reduction based on RG. Nonetheless, across many areas of science, simple models have been successful at capturing the salient features of complex systems, suggesting that a hierarchy of theories may exist even in the absence of a hierarchy of scales.

Recent work has attempted to formalize the emergence of simple effective theories in a broader context using methods from information geometry. The approach is grounded in the language of information theory and interprets the Fisher Information Matrix as a Riemannian metric on a parameterized space of models~\cite{Machta2013}. By studying the corresponding Riemannian manifold, termed the model manifold, the problem of model reduction is translated into the geometric problem of finding a lower-dimensional submanifold approximation to the model manifold~\cite{transtrum2014model}. Additionally, the information geometric approach has been used to study the loss of information under RG flow, showing analytically that, as coarse-graining proceeds, the metric deforms such that distances along relevant directions are preserved and distances along irrelevant directions are contracting~\cite{Raju2018}. 

Here we explore how effective theories emerge in systems without special symmetries and link the systematic elimination of degrees of freedom to the notion of a ``choice of question" rather than the notion of a hierarchy of scales. We show that the RG flow of model parameters can be replaced more generally by a flow of the Fisher Information Metric on the model manifold. Importantly, any theory becomes endowed with such a flow once the choice of question is specified; and this question-dependence is naturally encoded in the metric flow construction. We observe that, while any parameter flow induces a metric flow, a given metric flow cannot in general be realized by a parameter flow. The geometric implication is that only certain choices of question produce a metric flow that leaves the geometry of the model manifold invariant; in particular, the RG construction has this property. Our aim is to clarify the relation between RG and information geometry and propose a framework for studying the emergence of effective theories that pulls together threads from previous work.

\section{Encoding a question via a metric flow} \label{sec:flow}

\begin{figure}[t]
{\includegraphics[width = 0.9\linewidth]{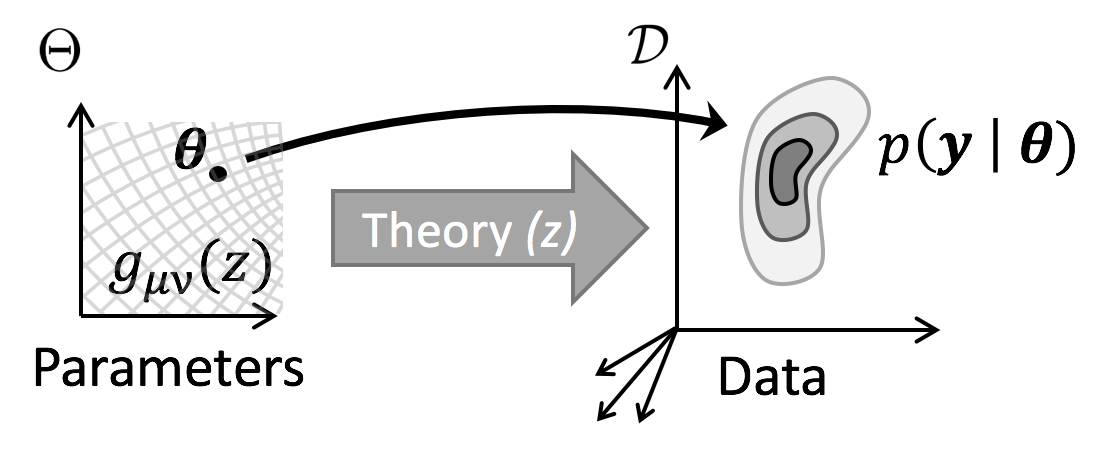}}
\caption{\textbf{Basic terms}
We define a theory as a mapping from a \emph{parameter space} $\Theta$ to probability distributions over the \emph{data space} $\mathcal{D}$. Each point $\v{\theta}\in \Theta$ defines a \textit{model}. The distinguishability of nearby models can be quantified by the Kullback-Leibler divergence between their predictions. This gives the Fisher Information metric on the parameter space $\Theta$, turning it into a Riemannian ``model manifold'' $\mathcal{M}$~\cite{Machta2013}. We extend this framework to consider smooth one-parameter deformations of the predictions, indexed by $z$. \label{fig:setup}
}
\end{figure}

A general way to conceptualize a physical theory is as a mapping from a parameter space $\Theta$ to probability distributions over a data space $\mathcal{D}$~\cite{mannakee2016sloppiness}. The parameter space spans the set of all possible values for each parameter and the data space comprises the corresponding predictions for measurable quantities of the theory. We use the term ``model'' to refer to individual points $\v{\theta} \in \Theta$. For a deterministic system with no measurement error, each point in $\Theta$ would map to a single point $\v{y} \in \mathcal{D}$, but in general the predictions of a given model will be probability distributions $p(\v{y} \,|\, \v{\theta})$ over the data space $ \mathcal D$ (Fig~\ref{fig:setup}A).

The sensitivity of model predictions to differential changes in parameters is quantified by the Fisher Information Metric (FIM), as in Ref.~\cite{Machta2013}: 
\begin{align}
&g_{\mu\nu}(\v{\theta})\, d\theta^\mu d\theta^\nu := D_{KL}
\left[p(\v{y} \,|\, \v{\theta}) \; || \; p(\v{y} \,|\, \v{\theta}+d\v{\theta})\right]
\nonumber\\
    & \qquad= \<\partial_\mu \log p(\v{y} \,|\, \v{\theta}) \;\partial_\nu \log p(\v{y} \,|\, \v{\theta})\>_p \, d\theta^\mu d\theta^\nu,
    \label{eq:FIM}
\end{align}
where $D_{KL}$ is the Kullback-Leibler divergence. 
This defines a distance metric on the parameter space, turning it into a Riemannian manifold, termed the model manifold $\mathcal{M}$. Note that it is sometimes useful to explicitly distinguish between $\mathcal M$ (as we defined it here) and its image under the theory mapping (i.e. its embedding into the space of probability distributions)~\cite{dufresne2016rigorSlopp}, in which case the term ``model manifold'' is reserved for the latter~\cite{transtrum2015perspective}; but in what follows, this distinction will not be important. Intuitively, the metric $g$ quantifies how distinguishable two models are based on their predictions. The global geometry of the model manifold encodes the overall structure of the theory and its possible reduced descriptions \cite{transtrum2014model,transtrum2016bridging}. This setup provides a formal context in which we can study the emergence of simpler effective theories.

\begin{figure*}[t]
{\includegraphics[width = 0.9\textwidth]{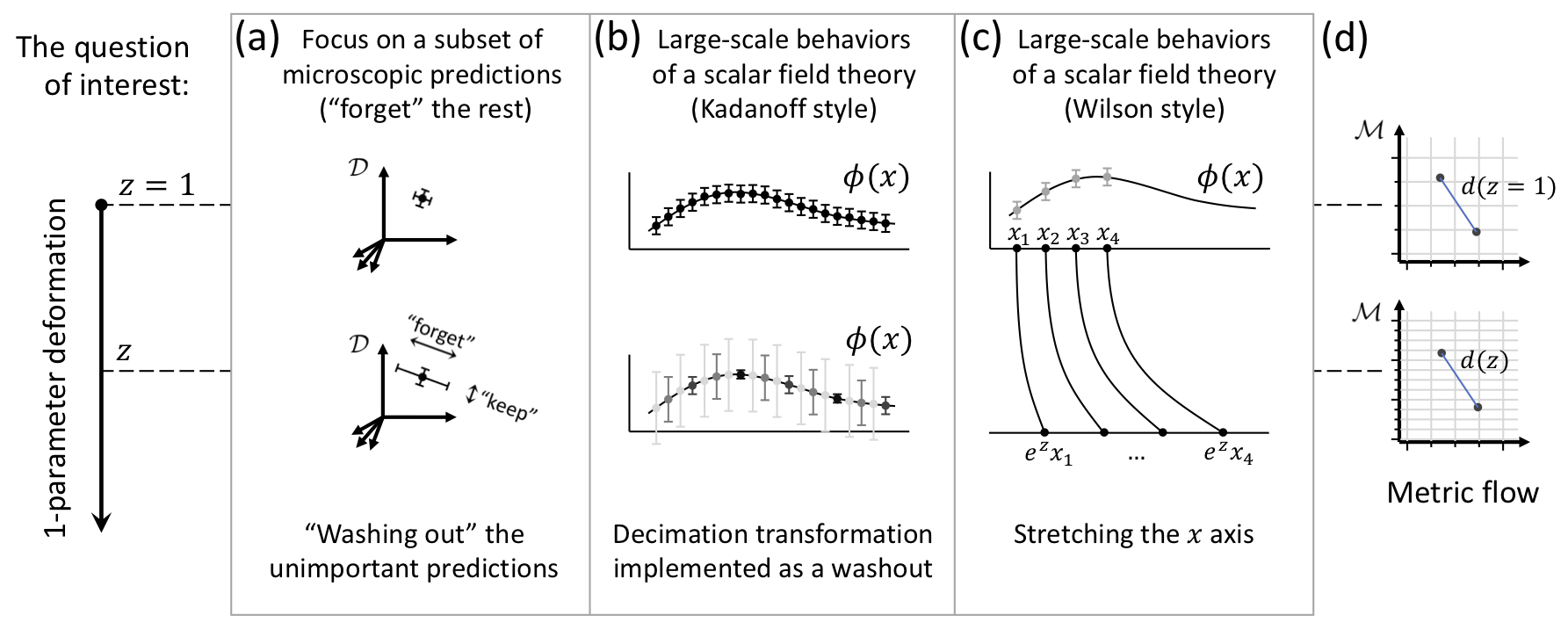}}
{\caption{\textbf{Implementing the question of interest as a smooth deformation of predictions,}
parametrized by $z$, as: $p(\v{y} \,|\, \v{\theta}) \to p(\v{y} \,|\, \v{\theta},z)$. (a-c) show three such example deformations. (a) illustrates the case where we start with some set of predictions and smoothly ``forget'' unimportant predictions by gradually increasing their measurement error. (b) shows how a similar procedure may be used to implement the ``block-spin'' real-space RG coarse graining: by increasing the error-bars on every 2nd spin exponentially faster than every 4th, etc. (c) implements RG with a very different 1-parameter deformation: by spreading out the points $x_i$ where the field $\phi(x)$ is being measured for the predictions $ p(\v{y} \,|\, \v{\theta},z)=p(\phi(\e{z}x_i)\,|\, \v{\theta}) $. All such examples induce an information metric flow $g_{\mu \nu}(z)$, wherein the way we measure distance on the model manifold $\mathcal{M}$ changes, as shown on the right (note: the tempting intuition of the manifold ``deforming'' may be misleading here, so we draw it as a fixed plane with changing definition of distances).
}\label{fig:question}}
\end{figure*}

Even when not stated explicitly, every effective
theory starts with choosing a question: deciding which properties of a system we care about and which we are willing to ignore. If we wanted to predict every microscopic detail of a system, no simplification would be possible; as noted by A. Rosenblueth and N. Wiener, ``the best material model of a cat is another, or preferably the same, cat'' ~\cite{wiener1945cats}. A general way to mathematically encode a ``choice of question'' is thus a necessary first step towards a unified method for constructing simple effective theories. Previous work, based on the Manifold Boundary Approximation Method (MBAM), has shown that defining different ``quantities of interest'' gives rise to different final reduced models~\cite{transtrum2014model,transtrum2016bridging}. Within MBAM, the boundaries of the model manifold are used to identify and remove parameter combinations, yielding a lower-dimensional submanifold. The boundaries represent physically interesting simplifying approximations, e.g. equilibrium, which are identified in the process.

Rather than projecting out the unimportant predictions in a discrete fashion, as in~\cite{transtrum2014model,transtrum2016bridging}, we here propose to consider a smooth construction to accommodate both reduction of multi-parameter models and RG coarse-graining within the same framework. 

If a theory is viewed as a mapping $\Theta \to P[\mathcal{D}]:\; \v{\theta} \to p(\v{y} \,|\, \v{\theta})$, we can introduce some smooth deformation of this mapping, parametrized by $z$, as $\theta \to p(\v{y} \,|\, \v{\theta}, z)$. We suggest that the space of such one-parameter deformations of the theory is general and flexible enough to encode any ``choice of question''. Fig.~\ref{fig:question} shows how this construct can encompass three very different questions: predicting a specific subset of the microscopic data, looking for large-scale system properties via Kadanoff real-space RG, and via Wilson-like RG, as discussed below. 

In all cases, such one-parameter deformation of the theory naturally induces a \textit{metric flow} $g_{\mu \nu}(z)$ via its definition in Eq.\ref{eq:FIM}. We argue that the question-dependent emergence of effective theories is encoded in the structure of this metric flow.

The first example in Fig.\ref{fig:question}(a) shows a typical model reduction, where we care about some predictions of the model and ignore the rest. For example, we may care about total protein abundance but not the individual protein copy numbers. Starting with the full set of predictions $p(\v{y} \,|\, \v{\theta})$, we smoothly wash out the unimportant data dimensions by gradually increasing their ``measurement error''. This may be implemented via a family of Gaussian convolution kernels $W_z(\v{y}) \sim \exp{-\frac{1}{2} y_i \,y_j \,w_{ij}(z)}$, such that the measured predictions transform as $p(\v{y} \,|\, \v{\theta},z) = p(\v{y} \,|\, \v{\theta}) * W_z(\v{y})$. Here $w_{ij}(z)$ controls the measurement error along various dimensions of $ \mathcal{D} $, and goes to $0 $ as $z \to \infty$ for the directions in data space we choose to ignore. As a result of this process, models that differ only in unimportant predictions gradually become indistinguishable, corresponding to a contraction of the model manifold $\mathcal{M}$.

For a theory where the predictions start out as deterministic $p(\v{y} \,|\, \v{\theta}) = \prod_i\delta(y_i - f_i(\v{\theta}))$, the ``wash-out protocol'' blurs each prediction with a Gaussian error and the metric flow assumes a particularly simple form:

\begin{align}
    g_{\mu\nu}(\v{\theta},z) = w_{ij}(z) \;\pd{f_i}{\theta^\mu} \pd{f_j}{\theta^\nu} \label{eq:wash-out}
\end{align}
The parameters of the original model that differentiated among the predictions we are choosing to ignore will correspond to the fastest-contracting directions in the model manifold $\mathcal{M}$. The remaining directions constitute the relevant parameter combinations for a simpler effective model. Note that the most/least relevant directions do not correspond to an eigendecomposition of the metric tensor itself, simply because eigenvectors can only be defined for linear transformations; diagonalizing a metric---a bilinear form---is mathematically ill-defined~\footnote{While the metric can of course be written out as a matrix in some particular basis, and the eigenvalues of that matrix can be calculated, the result will not be coordinate-invariant. In the present discussion,  the ``fastest-contracting directions'' are well-defined because the flow structure provides an object that can be contracted with the metric tensor to give a linear operator $\Lambda^\mu_\nu \equiv g^{\mu\lambda}\partial_z g_{\lambda\nu}$ with a nontrivial eigenbasis. Thus, implementing the question of interest via a washout construction captures the intuition of Ref.~\cite{Machta2013} in a covariant formalism.}. The characterization of systems in terms of the eigenvalues of the Fisher Information Metric in earlier work~\cite{Machta2013} generated some confusion, later clarified by the authors~\cite{transtrum2015perspective, dufresne2016rigorSlopp}. 

Focusing on large-scale behaviors of a theory is a more subtle notion than keeping only a subset of predictions; nevertheless, this kind of ``question of interest'' can be encoded in a similar manner.  Figure~\ref{fig:question}(b) illustrates a Kadanoff-style construction on a lattice, where the rate at which every 2nd lattice site is being washed out is set to be exponentially faster than every 4th site, and so on. This procedure is a smooth implementation of the sequential decimation used in real-space RG. Finally, the Wilson approach to RG also fits in the same framework, as shown in Figure~\ref{fig:question}(c) and described in detail below. Hence, encoding a question of interest via one-parameter deformations of a theory allows us to consider reduction of multi-parameter models and RG coarse-graining within the same framework.

\section{Metric flow vs parameter flow} \label{sec:RG}

The metric flow and the RG flow are qualitatively different mathematical objects. The RG flow is a point flow of parameters $ \v{\theta}(z) \in \Theta $; and via the correspondence $ \v{\theta} \to p(\v{y} \,|\, \v{\theta})$, we can similarly say that it maps points of $\mathcal{M}$ to other points of $\mathcal{M}$. In contrast, as illustrated in Fig.~\ref{fig:question}(d), a metric flow merely changes how we measure distance between points (which do not move) in $\mathcal{M}$; the manifold itself does not deform~\footnote{However, since manifolds are typically visualized by picking an embedding into a larger space~\cite{quinn2017}, visualizing a manifold endowed with a metric flow will appear to deform it. It is therefore worth emphasizing that this deformation would then be a deformation of the \emph{embedding}, and not a point flow on the manifold.}.

\begin{figure}[t]
\centering
\includegraphics[width=\linewidth]{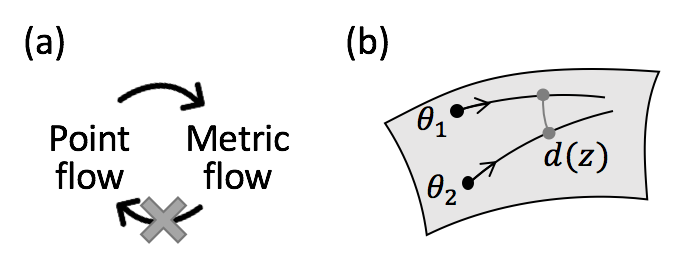}
\caption{\textbf{The metric flow construction is more general than the point flow arising in RG-treatment.}
(a) Any point flow induces a metric flow, by a standard mathematical construction called pullback (along a vector field), as in Eq.\ref{eq:Lie_flow}. However, a metric flow cannot in general be realized by a point flow.
(b) For renormalizable theories, the point flow of the renormalization group induces precisely the metric flow discussed here.
} 
\label{fig:pointFlow}
\end{figure}

Formally, we can see that any point flow on a manifold, including the RG flow, induces a metric flow simply by following the stream-lines:
\begin{align}
    \partial_z `g_mn` = \mathcal{L}_{\vec{\beta}}\, `g_mn`,
    \label{eq:Lie_flow}
\end{align}
where $\mathcal{L}_{\vec{\beta}}$ is the Lie derivative along a vector field $\vec{\beta}$ (a directional derivative of a tensor), see Fig.~\ref{fig:pointFlow}(b). However, the converse is not true: for a given metric flow, there is no general way to construct a corresponding point flow field; e.g. a uniform rescaling of the canonical metric on a sphere cannot be realized as a flow of points on that sphere. (More generally, if $d$ is the number of model parameters, then equation~\eqref{eq:Lie_flow}, interpreted as an equation for the vector field $\vec{\beta}$, has $\bigO{d^2}$ constraints for $\bigO{d}$ unknowns.) The metric flow can be defined for any system and question; in contrast, realizing it as a point flow requires a special structure. Systems for which we can apply the RG procedure are in this special class. Raju et al. \cite{Raju2018} demonstrated that, under RG flow, the metric deformation preserves distances along relevant directions, while the distances along irrelevant directions contract according to the RG exponents. Hence, the RG point flow induces the expected metric flow, corresponding conceptually to the top arrow in Fig.~\ref{fig:pointFlow}(a). 

The fact that the metric flow induced by the RG point flow according to Eq.~\eqref{eq:Lie_flow} is identical to the metric flow generated by a one-parameter deformation $p(\v{y} \,|\, \v{\theta}, z)$ that implements a Wilson-style coarse-graining can be shown in all generality. The appropriate one-parameter deformation here is changing the observation length-scale: if our system is a field theory, and predictions describe the field values at certain points $y_i = \phi(x_i)$, then the deformed predictions would be given by $p(\v{y} \,|\, \v{\theta},z) = p(\{\phi(\e{z}x_i)\} \,|\, \v{\theta})$. We can think of this as measuring the field with a lattice of detectors, which we spread apart as $ z $ increases [Fig.~\ref{fig:question}(c)]. Equation~\eqref{eq:FIM} then gives the corresponding metric flow $ g(z) $. Since the RG flow $ \v{\theta}(z) $ by definition gives the parameters that describe the system at different scales, we have that $ p(\phi(\e{z}x_i)\,|\, \v{\theta}(0))=p(\phi(x_i)\,|\, \v{\theta}(z)) $. This allows us to write
\begin{align} \label{eq:metricLieFlow}
\partial_z `g_mn` &= \partial_z \<\partial_\mu \log p(\phi| \v{\theta}(z)) \;\partial_\nu \log p(\phi| \v{\theta}(z))\>_\phi \nonumber\\
&=\<\partial_\mu \left[\dot{\theta}^\lambda\, \partial_\lambda\log p(\phi| \v{\theta}(z))\right] \;\partial_\nu \log p(\phi|\v{\theta}(z))\>_\phi \nonumber\\ 
& \hspace{0.6\linewidth}+ (\mu \leftrightarrow \nu) \nonumber\\
&=\mathcal{L}_{\dot{{\theta}}}\, `g_mn`,
\end{align}
where in the second line we commute the $ z $ derivative with the field average, and in the third line we expand the $ \theta_\mu $ derivative and use the component-wise definition of the Lie derivative:
$
    \mathcal{L}_{\vec{\beta}}\, `g_mn` \equiv `\beta^l``\partial_l``g_mn` + `g_ml``\partial_n``\beta^l` + `g_ln``\partial_m``\beta^l`
$. The fact that we are able to recover Eq.~\eqref{eq:Lie_flow} verifies that the metric flow due to the deformation of the theory is indeed the same as that induced by the parameter flow in the case of RG\footnote{In ~\cite{Raju2018}, Raju et al. use a modified Lie derivative containing an additional term that describes the change in the metric due to the shrinking of the effective system size upon coarsegraining. In our construction, the effective system size, and therefore the total amount of data, is preserved as the scale of observation increases.}. Consequently, the theories traditionally studied with RG comprise special cases where the metric flow is realizable by a point flow. 

\section{A geometric perspective} \label{sec:augM}

\begin{figure}[t]
\centering
\includegraphics[width=\linewidth]{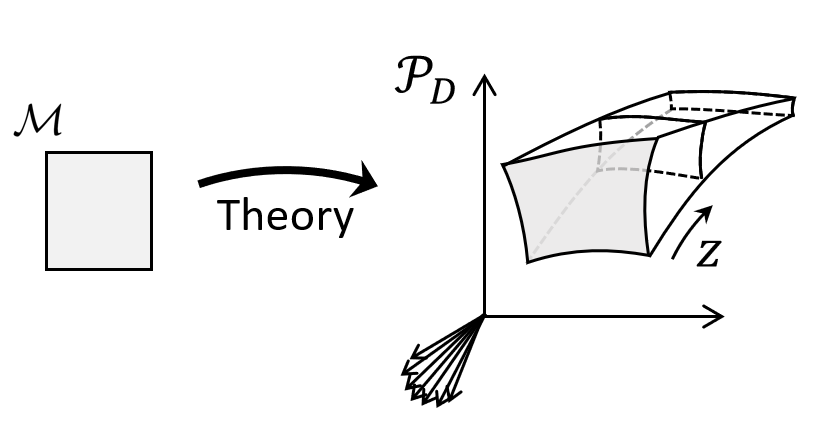}
\caption{\textbf{The augmented manifold.}
Consider the embedding of the model manifold $ \mathcal{M} $ into the space of all probability distributions over data space (this is an incredibly vast space). In this embedding, a general one-parameter deformation of the theory will cause the model manifold to move off its original embedding as $z$ increases. We construct the augmented model manifold by including the deformation parameter $z$ as a new parameter in $\mathcal{M}$ and considering the embedding of the resulting $(d+1)$-dimensional parameter space. In the special case of RG, this new dimension $ z $ will be degenerate in the augmented manifold, which will consequently have zero thickness along this direction.} 
\label{fig:augm}
\end{figure}

We have shown that only in special cases can the metric flow be realized by a point flow and that, in particular, the RG construction has this property. But what are the geometric implications of this in terms of the model manifold $\mathcal{M}$ and what does that tell us about which systems might be amenable to an RG-like approach?

The self-similarity across scales that is required for RG ensures that the coarse-grained system can be described by a theory of the same form as the original system, just with a different set of parameters. Hence the point flow constitutes a relabelling of points on the model manifold (i.e., a change of coordinates), and by construction leaves any coordinate-invariant geometric and topological features of the manifold fixed. More formally, we can say that the family of metrics in RG-induced metric flow are all diffeomorphic to each other. This need not be the case for more general metric flows, where all features of the manifold can deform under the flow. 

One way to visualize this is to consider the embedding of the model manifold $ \mathcal{M} $ into the space of all probability distributions over data space (this is an incredibly vast space). In this embedding, a general one-parameter deformation of the theory will cause the model manifold to move off its original embedding as $z$ increases, see figure \ref{fig:augm}. We can formalize this by constructing an \textit{augmented model manifold} by including the deformation parameter $z$ as a new parameter in $\mathcal{M}$. This way, instead of the metric flow $ g_{\mu\nu}(\v{\theta},z) $, we now consider a $ (d+1) $-dimensional augmented Fisher Information Metric $ G_{\tilde{\mu}\tilde{\nu}} (\tilde{\v{\theta}})$, where $ \tilde{\v{\theta}} = \left(\left\{\theta_\mu\right\}, z\right) $. In the case of RG, the augmented manifold is degenerate: the augmented metric $ G(\tilde{\v{\theta}}) $ is null along the RG flow lines and the manifold will consequently have zero ``thickness'' in the $ z $ dimension.

The degeneracy of the augmented manifold is a defining geometric property of the RG procedure~\footnote{Technically, the augmented manifold being degenerate (zero-width) in the $z$-direction is a geometric criterion that could, in principle, be satisfied by a theory without a strict self-similarity symmetry. Such a theory would then admit the full analog of an RG construction, but we have not identified a nontrivial example.}. By construction, in RG we start with the most general Hamiltonian allowed by symmetry, including parameters for non-local couplings that are zero in the original Hamiltonian but become non-zero as coarse-graining proceeds~\cite{goldenfeld2018lectures}. Since the symmetries of the system are preserved by the RG coarse-graining, the flow is guaranteed to stay within the original embedding of the model manifold. If we were to leave out any relevant parameters when defining the parameter space, those directions would be generated under coarse-graining $z$, and the augmented manifold would be extended along that direction. It is intriguing to speculate whether cases where the thickness of the augmented manifold in the $z$-direction is non-zero but small could allow defining a useful notion of ``RG-like" systems.

\section{Summary}
We presented a framework that clarifies the mathematical connection, and more importantly the distinction, between the ideas developed from information geometry for studying how simple effective theories emerge and the canonical RG construction for coarse-graining. We showed how various ways to ``specify a question'' can be implemented in a unified way as a 1-parameter deformation of the theory. We then discussed how the induced metric flow differs from a parameter flow familiar from RG and showed how the two can be reconciled using an augmented manifold construction. Although the renormalization group is the canonical construction for studying the emergence of effective low-dimensional theories, we expect that generalizing this understanding to theories without special symmetries will operate with the concept and properties of a metric flow.

\acknowledgements{We thank Benjamin Machta and Mark Transtrum for helpful discussions. PC was supported in part by the James S. McDonnell Foundation Scholar Grant 220020476, and by the International Centre for Theoretical Sciences (ICTS) during a visit for participating in the program -US-India Advanced Studies Institute: Classical and Quantum Information (Code: ICTS/Prog-infoasi/2016/12).}

\nocite{*}
\bibliography{References}

\end{document}